\documentclass[conference]{IEEEtran}
\IEEEoverridecommandlockouts

\makeatletter
\def\ps@headings{%
\def\@oddhead{\mbox{}\scriptsize\rightmark \hfil \thepage}%
\def\@evenhead{\scriptsize\thepage \hfil \leftmark\mbox{}}%
\def\@oddfoot{}%
\def\@evenfoot{}}
\makeatother
\usepackage{ifpdf}

\usepackage{cite}

\ifCLASSINFOpdf
  \usepackage[pdftex]{graphicx}
\else
  \usepackage[dvips]{graphicx}
\fi
\usepackage[cmex10]{amsmath}
\usepackage{algorithmic}

\usepackage{array}

\usepackage{mdwmath}
\usepackage{mdwtab}

\usepackage{eqparbox}

\usepackage[font=footnotesize]{subfig}
\usepackage{fixltx2e}

\usepackage{stfloats}

\usepackage{url}

\usepackage{color}

\newtheorem{lem}{Lemma}

\newtheorem{thm}{Theorem}

\newtheorem{prop}{Proposition}

\begin{document}
%
\title{Comparative Statics On The Allocation Of Spectrum}

\author{\IEEEauthorblockN{Vijay G. Subramanian, Mike Honig, Randy Berry}
\IEEEauthorblockA{
EECS Department\\ Northwestern University,
Evanston, IL 60208 \\
Email: v-subramanian@northwestern.edu, [mh, rberry]@eecs.northwestern.edu}
}


%


\maketitle

\begin{abstract}
Allocation of spectrum is an important policy issue and decisions taken have ramifications for future growth of wireless communications and achieving universal connectivity. In this paper, on a common footing we compare the social welfare obtained from the allocation of new spectrum under different alternatives: to licensed providers in monopolistic, oligopolistic and perfectly competitive settings, and for unlicensed access. For this purpose we use mathematical models of competition in congestible resources. Initially we assume that any new bandwidth is available for free, but we also generalize our results to include investment decisions when prices are charged for bandwidth acquisition.
\end{abstract}


\section{Introduction}\label{sec:intro}

Spectrum allocation and its impact on commercial wireless mobile broadband as well as the goal of universal electronic connectivity has been a topic of great interest in recent years. A large driver for this has been the take-up and usage of mobile telephony and wireless-data-capable devices. This has lead to increased demand from the wireless service providers for spectrum~\cite{CNNMoney2012, CT2012, NYT2012}. At the same time there have been many scientific articles~\cite{Lazarus2010} on current spectrum usage and the impact of policy decisions on the current wireless broadband environment~\cite{Hazlett2005, Hazlett2006, Benkler2012}. The issues and concerns raised have lead to further policy discussions~\cite{FCC2010BP, PCAST2010} on freeing up more spectrum and also on the means of providing this for the purposes of wireless broadband access. One proposal has been to auction the freed up spectrum to commercial wireless service providers~\cite{FCC2012}. Another proposal that builds on the success of open-access strategy~\cite{Benkler1998, Benkler2002} of 802.11/WiFi~\cite{IEEE1999} for the 2.4 GHz and 5GHz bands, is to provision~\cite{WeiserHatfield2005, EilatLevinMilgrom2011} the new spectrum also for open/unlicensed access, also known as spectrum commons. A step along this direction has already been taken by the FCC for the TV white-space~\cite{FCC2010WS}. In the rest of the document we will use commons, white-space, open-access and unlicensed access interchangeably. 

There has been considerable debate on these different spectrum allocation proposals~\cite{Hazlett2005, Hazlett2006, NZBHV_D2011, NZBHV_N2011, EilatLevinMilgrom2011, Benkler2012}, both on technical issues and on their impact on social welfare. It has been argued in \cite{Hazlett2005, Hazlett2006} that following Coase's theorem~\cite{Coase1959, Coase1960}, the best approach is to properly define the property rights of owing spectrum so that commercial entities can then trade purchased (or allocated) spectrum among each other in order to achieve a socially optimal or efficient outcome. On the contrary, arguing that the open-access philosophy of 802.11/WiFi spurred multiple innovations in wireless technologies and devices leading to greater social welfare over time, \cite{Benkler1998, Benkler2002, EilatLevinMilgrom2011, Benkler2012} have advocated for a large unlicensed access component of any new release of spectrum. The work in \cite{NZBHV_D2011, NZBHV_N2011} advocated caution in adopting the commons paradigm with the prevalent spectrum allocations. They analyzed markets where unlicensed spectrum coexists with licensed spectrum and showed the both social welfare and consumer welfare could decrease when new spectrum is added as unlicensed spectrum.

The purpose of this article is to consider a simple analytic model where the different options of allocating bandwidth described above can be compared in quantitative terms while taking into account existing bandwidth allocations and market structures in terms of commercial providers and unlicensed access. In particular, we would like to compare the following market scenarios in terms of their impact on total social welfare and consumer surplus when additional bandwidth is provisioned: monopoly, monopoly with unlicensed access, oligopoly, oligopoly with unlicensed access, perfect competition (where a large number of providers is considered) and perfect competition with unlicensed access. Initially, we develop our results assuming that the additional bandwidth is available free of charge. However, we then use these results to the setting where commercial entities have to purchase bandwidth. As a consequence, we propose per unit price valuations than can be used to compare different market scenarios.

The paper is structured as follows. In Section~\ref{sec:model}, we describe the model of competition in congestible resources that we use for all our results, also set-up the require notation. Section~\ref{sec:results} is devoted to characterizing the equilibria when the capacity is assumed to be given free of charge. In Section~\ref{sec:invest}, we use the results developed in Section~\ref{sec:results} to include investment decisions by the providers when they have to pay to obtain spectrum. We numerically explore the consequences of our results in Section~\ref{sec:discuss}, and conclude in Section~\ref{sec:conc}.

\section{Model}\label{sec:model}

We use the model of competition in congestible resources~\cite{OzdaglarSrikant2007, Ozdaglar2008, HuangOzdaglarAcemoglu2006, AcemogluOzdaglar2007, RoughgardenTardos2004, Roughgarden, HayrapetyanTardosWexler2005, BaakeMitusch2007, JohariWeintraubVanRoy2010, AcemogluBimpikisOzdaglar2009, NZBHV_D2011, NZBHV_N2011, AAW_Perf2011} built upon the use of the Wardrop equilibrium~\cite{Pigou1920, Wardrop1952}. The main tenet is that impact of power-sharing and interference constraints in wireless systems on the achieved quality of service is conceptualized as latency prices that get added to any service price for the consumers. In addition, when faced with alternatives for access/usage, the users differentiate based on total delivered price, which is the sum of access price and latency price, and choose to use only those that have the lowest price; this is the Wardrop equilibrium~\cite{Wardrop1952} condition. In the mathematical model, users are assumed to be atomic, i.e., a mass of consumers in $[0,q_{\max}]$, and have a concave decreasing demand-to-price function $P(q):[0, q_{\max}]\mapsto[0,p_{\max}]$ for quantity $q$, with inverse $Q(p):[0, p_{\max}]\mapsto[0,q_{\max}]$ for price $p$; note that $Q(p)$ is also decreasing but convex. The wireless systems are assumed to have a latency function for spectrum usage level $q$ given by $l(q/C)$ that is convex increasing with $l(0)=0$, where $C$ is the provisioned capacity. For a usage level of $q$, provisioned capacity $C$ and asking price $p$, the delivered price is $p+l(q/C)$ and consumers compare different access possibilities using this delivered price. 

Since we're comparing different types of systems, we will index the quantity/demand served $q$, the prices assessed $p$, the capacities provisioned and the latency functions as follows: by sub-script $m$ if a monopoly, $1$ and $2$ to indicate competitive providers for the oligopolistic scenario, $w$ for white-space/open-access, and $c$ for perfect competition (i.e., atomic service providers); for the oligopolistic scenario we will also denote the capacities by the appropriate indices. The capacity allocated to white-space/open-access will be denoted by $W$, and following the results of \cite{NZBHV_D2011, NZBHV_N2011}, we will assume that the price assessed for whitespace/open-access is $0$ so that any service providers operating in these bands will not earn any revenue for users served; note that there will still be the latency charge based on usage level which the users will perceive and react to. In general, we will assume that $P(\cdot)$, $l(\cdot)$ and $D(\cdot)$ are all twice continuously differentiable. Many of our results and calculations will be for linear latency functions and demand functions, i.e., we will set $l(x)=x$, $P(q)=p_{\max}(1-q)$ (assuming $q_{\max}=1$ without loss of generality) and $D(p)=1-p/p_{\max}$. In many cases we will make the simplifying assumption that the latency function for all the entities involved is the same, i.e., some function $f(\cdot)$. For further details, we refer the readers to \cite{AcemogluOzdaglar2007, HayrapetyanTardosWexler2005, AcemogluBimpikisOzdaglar2009, JohariWeintraubVanRoy2010, NZBHV_D2011, NZBHV_N2011, AAW_Perf2011}, and also to \cite{SMR_2013}.

\section{Equilibria given initial endowments}\label{sec:results}

We start by considering a monopoly provider setting. Here the provider sets a price for access and the user population demand is determined by the total delivered price, i.e., the sum of the access price and the latency price. In this setting, the optimal price for the monopolist is one that maximizes her revenue. Under the assumptions of our model, the revenue optimal price is unique. Furthermore, adding additional bandwidth to the monopolist's portfolio results in an increase in total welfare. These results are summarized below.
\begin{prop}\label{prop:M}
With a monopoly service provider, the revenue maximization problem is a convex optimization problem where the user equilibrium demand satisfies the following fixed point equation
\begin{align}
q_m^*= - \frac{P(q_m^*)-l_m\left(\frac{q_m^*}{C}\right)}{P^\prime(q_m^*) - \frac{l_m^\prime\left(\frac{q_m^*}{C}\right)}{C}}
\end{align}
with a unique solution in $[0, q_{\max}]$. Given the user equilibrium demand, the change in total welfare with additional capacity allocated to the monopolist is non-negative and given by
\begin{align}
\frac{\partial T_m(q_m^*)}{\partial C} = - q_m^* P^\prime(q_m^*) \frac{\partial q_m^*}{\partial C} + q_m^* \frac{l_m^\prime\left(\frac{q_m^*}{C}\right)}{C^2}.
\end{align}
\end{prop}
\begin{IEEEproof}
See Appendix~\ref{sec:monopoly}.
\end{IEEEproof}
With a linear latency function and a linear demand to price function, the user equilibrium demand, the monopolist price, the revenue, consumer surplus and change in total welfare with additional capacity are explicitly given by
\begin{align*}
& q_m^*  =\frac{1}{2} \frac{C p_{\max}}{1+C p_{\max}}, \quad S_m(q_m^*)  = \frac{p_{\max}}{8} \left( \frac{C p_{\max}}{1+C p_{\max}} \right)^2, \\
& p_m^*  = \frac{p_{\max}}{2}, \quad R_m(q_m^*)  = \frac{p_{\max}}{4} \frac{C p_{\max}}{1+C p_{\max}}, \\
& \frac{\partial T_m(q_m^*)}{\partial C} = \frac{p_{\max}}{4 (1 + C p_{\max})^2} \left( 1 + \frac{C p_{\max}}{1+C p_{\max}} \right).
\end{align*}


Next we consider a scenario where the monopolist provider has to compete with an open-access band. We add $W$ capacity as whitespace and consider the impact of this. We again define $q_c$ to be the quantity served in the proprietary spectrum and set $q_w$ to be the quantity served in the whitespace. As mentioned before, we will sometimes make the simplifying assumption that $l_m(q_m/C)=f(q_m/C)$ and the latency cost in the whitespace $l_w(q_w/W) = f(q_w/W)$ where $f(\cdot)$ is an increasing convex function with domain being the non-negative reals.

We start by assuming that there is no proprietary spectrum. Then quantity served in the whitespace is given by the unique solution to
\begin{align*}
l_w\left( \frac{q_w}{W}\right) = P(q_w).
\end{align*}
Denote the solution by $\hat{q}_w$. Note that if the proprietary provider sets an asking price greater than $P(\hat{q}_w)$, then no traffic arrives, and such a strategy will not be adopted as a revenue maximizing solution. For an asking price less than $P(\hat{q}_w)$, there will be some traffic to the proprietary provider's spectrum such that a user equilibrium is achieved, i.e., the delivered price from both spectrum sources will be the same so that users are indifferent; that the asking price is less than $P(\hat{q}_w)$ will be a standing assumption in the rest of this document. This then implies that
\begin{align*}
p + l_m\left( \frac{q_m}{C}\right) = l_w\left( \frac{q_w}{W}\right) = P(q_m + q_w).
\end{align*}
The revenue of the proprietary provider, consumer surplus and total welfare are then given by
\begin{align*}
R_{mw}(p,q_m,q_w) & = p q_m \\
S_{mw}(p, q_m, q_w) 
& = \int_0^{q_m + q_w} P(q) dq - (q_m+q_w) P(q_m + q_w) \\
T_{mw}(p,q_m,q_w) & = \int_0^{q_m+q_w} P(q) dq - q_m l_m\left(\frac{q_m}{C}\right) - q_w l_w\left(\frac{q_w}{W}\right)
\end{align*}
where we have used the user equilibrium relationships to derive the alternate expressions. We use the subscript ``mw" to denote the monopoly with whitespace scenario.

From the user equilibrium we get
\begin{align*}
q_m & = D\left( l_w \left(\frac{q_w}{W}\right) \right) - q_w \\
p & = l_w\left( \frac{q_w}{W}\right) - l_m\left( \frac{D\left( l_w\left(\frac{q_w}{W}\right) \right) - q_w}{C}\right).
\end{align*}
The above relationship explicitly characterizes the fact that there is only one degree of freedom. For ease of analysis we will consider $q_w$ to be the degree of freedom. Constraining the range of $p$ to be $[0, P(\hat{q}_w)]$ implies a similar constraint on $q_w$ (and $q_m$). One bound on $q_w$ is $\hat{q}_w$. The second is given by the case when $p=0$ where we have
\begin{align*}
l_m\left( \frac{q_m}{C}\right) = l_w\left( \frac{q_w}{W}\right) = P(q_m + q_w).
\end{align*}
There it is easy to see that solutions are $\tilde{q}_w$ and $\tilde{q}_m=C\; l_m^{-1}( l_w(\tilde{q}_W / W))$ where $\tilde{q}_w$ is the unique solution to 
\begin{align*}
l_w\left( \frac{q_w}{W}\right) = P\left(q_w+C\; l_m^{-1}\left( l_w\left(\frac{\tilde{q}_W}{W}\right)\right)\right)
\end{align*}
It is easy to see that $\tilde{q}_w < \hat{q}_w$ so that the allowed values of $q_w \in [\tilde{q}_w, \hat{q}_w]$. With these definitions, the revenue maximization problem is given by
\begin{align*}
& \max_{q_w \in [\tilde{q}_w, \hat{q}_w]}  R_{mw}(p,q_c,q_w)=p q_m 
\end{align*}

In \cite{NZBHV_D2011, NZBHV_N2011} the authors showed for box demand functions and linear latency functions that the total welfare can decrease upon the introduction of white-space. 
This reduction in total welfare with the addition of a small amount of white-space is even more general. For general latency functions and general demand versus price functions, we have the following result.
\begin{thm}\label{thm:MW}
Let $q_m^*$ and $p_m^*$ be the user equilibrium demand and the revenue maximizing price in the monopoly setting. Using these define
\begin{align*}
 \hat{q}_w   = l_w^{-1}\left(p_m^*+l_m\left( \frac{q_m^*}{C}\right) \right) = l_w^{-1}(P(q_m^*))
\end{align*}
to obtain the marginal demand served in the white-space.
Then the change of total welfare with the addition of a small amount of white-space is given by
\begin{align}\label{eq:Tmw}
\begin{split}
& \frac{\partial T_{mw}}{\partial W}\Big|_{W=0}  =  \\
& \frac{1}{2} \frac{\left(\hat{q}_w +  q_m^*\frac{P^\prime(q_m^*)}{l_w^\prime(\hat{q}_w)}\right) P^\prime(q^*_m) + q_m^* \hat{q}_w P^{\prime\prime}(q^*_m)}{P^\prime(q^*_m)  - \frac{l_m^\prime\left( \frac{q_m^*}{C}\right)}{C}+  q_m^* \frac{P^{\prime\prime}(q^*_m)}{2}} q_m^* P^\prime(q_m^*).
\end{split}
\end{align}
The marginal change in total welfare with the addition of a small amount of white-space is negative when either $\hat{q}_w +  q_m^*\tfrac{P^\prime(q_m^*)}{l_w^\prime(\hat{q}_w)}\geq 0$ or the latency functions are linear. 
\end{thm}
\begin{IEEEproof}
See Appendix~\ref{sec:monopolyplus}.
\end{IEEEproof}
When the demand functions are linear and $l_m(x)=l_w(x)=x$, the revenue maximization problem is a convex optimization problem with a unique solution; we omit the details as the analysis is similar to the proof of Proposition~\ref{prop:M}. In this case the user equilibrium demands, the monopolist price, the optimal revenue, consumer surplus and the change in total welfare with additional capacity are explicitly given by
\begin{align*}
& q_w^* = \frac{\frac{C p_{\max}}{2} + W p_{\max} + 1}{C p_{\max}+ W p_{\max} + 1} \frac{W p_{\max}}{W p_{\max} + 1}, \\
& q_m^*  = \frac{1}{2} \frac{C p_{\max}}{C p_{\max}+ W p_{\max} + 1}, \quad p_m^*  = \frac{p_{\max}}{2} \frac{1}{1+W p_{\max}}, \\
& R_{mw}(q_w^*)  = \frac{p_{\max}}{4} \frac{C p_{\max}}{C p_{\max}+ W p_{\max} + 1}\frac{1}{1+W p_{\max}}, \\
& S_{mw}(q_w^*)  = \frac{p_{\max}}{2} \left(\frac{W p_{\max} + C p_{\max} - \frac{C p_{\max}}{2 (1 + W p_{\max})}}{C p_{\max}+ W p_{\max} + 1}\right)^2, \\
& \frac{\partial T_{mw}(q_w^*)}{\partial W}\Big|_{W=0}    = -\frac{p_{\max}^2}{4} \frac{C^2 p_{\max}^2}{(1+C p_{\max})^3},
\end{align*}
We will be using this to compare across the different market types.

Next we consider a competitive setting with two providers. We will use the results of this scenario to determine the impact of creating a competitor by assigning the new bandwidth to a different entity. In the two providers competitive environment, the work in \cite{NZBHV_D2011, NZBHV_N2011} observed that allocating the additional spectrum as licensed provides higher social value with box demand functions and linear latency functions. 
In order to compare across market types in a uniform manner and to obtain explicit expressions, we will, however, consider linear latencies and linear demand versus price functions; we will also assume that the latency functions are the same for both providers. The existence of equilibria can be proved under more general conditions following the approach in \cite{AcemogluOzdaglar2007, AcemogluBimpikisOzdaglar2009, JohariWeintraubVanRoy2010}. We let provider $i$ have capacity $C_i$ and revenue $R_i$ where $i\in\{1,2\}$. Note that we have a game formulation between the two providers that is based upon the user equilibrium for prices chosen by the providers. The resulting sub-game perfect equilibrium is summarized below.
\begin{prop}\label{prop:Nc}
The oligopoly model of spectrum access admits a unique sub-game perfect equilibrium. The unique sub-game perfect equilibrium of prices set by competitor $i=1, 2$ is given by
\begin{align}\label{eq:Ncp}
p_i^* & = p_{\max} \frac{p_{\max}^2 + \frac{p_{\max}}{C_i}+2 \frac{p_{\max}}{C_{-i}}  + \frac{2}{C_1 C_2}}{3 p_{\max}^2 + 4 p_{\max} \left(\frac{1}{C_1}+\frac{1}{C_2} \right) + \frac{4}{C_1 C_2}} 
\end{align}
The corresponding user equilibrium demands for $i=1, 2$ are
\begin{align}\label{eq:Ncq}
q_i^* & = \frac{p_{\max}^2 + 2 \frac{p_{\max}}{C_{-i}}}{3 p_{\max}^2 + 4 p_{\max} \left(\frac{1}{C_1}+\frac{1}{C_2} \right) + \frac{4}{C_1 C_2}}
\end{align}
\end{prop}
\begin{IEEEproof}
See Appendix~\ref{sec:oligopoly}.
\end{IEEEproof}
Using the expressions in \eqref{eq:Ncp} and \eqref{eq:Ncq}, the revenues of the providers, the consumer surplus and the total surplus are given by
\begin{align*}
R_1(p_1^*,p_2^*) & = p_1^* q_1^*, \quad R_2(p_2^*,p_1^*)  = p_2^* q_2^*, \\
S_o(p_1^*,p_2^*) & = \frac{p_{max}}{2} (q_1^*+q_2^*)^2, \\
T_o(p_1^*,p_2^*) & = R_1(p_1^*,p_2^*) + R_2(p_2^*,p_1^*) + S_o(p_1^*,p_2^*).
\end{align*}
We use the subscript ``o" to denote oligopolistic competition.

Next consider the scenario of two competitors with white-space; again assume that the latency function is the same for white-space. We can extend the result of Proposition~\ref{prop:Nc} to the following characterization of the sub-game perfect equilibrium. Define $D$ to be the following quantity
\begin{align}
\begin{split}
D& :=3 p_{\max}^2 + 4 p_{\max} \left(\frac{1}{C_1}+\frac{1}{C_2} \right)\\
& \qquad +4W p_{\max}^2 \left( 2+ 2p_{\max} + \frac{1}{C_1}+\frac{1}{C_2}\right)\\
& \qquad +4 \prod_{i=1}^2 \left(\frac{1}{C_i} + W p_{\max} \left( 1+ p_{\max} + \frac{1}{C_i}\right)\right) 
\end{split}
\end{align}
\begin{prop}\label{prop:Ncw}
The oligopoly model of spectrum access with additional open-access admits a unique sub-game perfect equilibrium. The unique sub-game perfect equilibrium of prices set by competitor $i=1, 2$ is given by
\begin{align}
\begin{split}
& \frac{p_i^*}{p_{\max}}  =  \left( 2 W p_{\max} + W p_{\max}^2 +(1+W p_{\max}) \left(p_{\max} + \frac{1}{C_i}\right)\right) \\
& \quad \times \frac{\left( 2 W p_{\max} + W p_{\max}^2 +(1+W p_{\max}) \left(p_{\max} + \frac{2}{C_{-i}}\right)\right)}{(1+W p_{\max}) D} 
\end{split}
\end{align}
The user equilibrium demands are given by
\begin{align}
\begin{split}
& \forall i=1, 2 \;\; q_i^*  = \frac{2 \frac{p_{\max}}{C_{-i}} + p_{\max}^2 +2 W p_{\max}^2 \left( 1+ p_{\max} + \frac{1}{C_{-i}}\right)}{D} \\
& q_w^*  = \frac{W p_{\max}}{1+W p_{\max}} \\
& \qquad \times \frac{\prod_{i=1}^2\left(p_{\max} +\frac{2}{C_i}+2 W p_{\max} \left(1+p_{\max} +\frac{1}{C_i} \right)\right)}{D} 
\end{split}
\end{align}
\end{prop}
\begin{IEEEproof}
See Appendix~\ref{sec:oligopolyplus}.
\end{IEEEproof}
The revenues of the providers, the consumer surplus and the total surplus are given by
\begin{align}
\begin{split}
R_1(q_1^*,q_2^*) & =p_1^* q_1^*, \quad R_2(q_2^*,q_1^*)  =p_2^* q_2^*,\\
S_{ow}(q_1^*,q_2^*) & = \frac{p_{\max}}{2} (q_1^*+q_2^*+q_w^*)^2, \\
T_{ow}(q_1^*,q_2^*) & = R_1(q_1^*,q_2^*) + R_2(q_2^*,q_1^*) + S_{ow}(q_1^*,q_2^*).
\end{split}
\end{align}
Here we use the subscript ``ow" to denote the oligopolistic competition with whitespace scenario.

Now consider perfect competition, i.e., the limiting regime of infinitely many providers. The applicability of the atomic providers model of perfect competition can be formally justified as a limit of (symmetric) Nash equilibria of the symmetric finite provider model (sub-case of the model in Prop.~\ref{prop:Nc} when there are two providers) as the number of providers increases without bound; the details can be found in Appendix~\ref{sec:perfectcompetition}, 
also see \cite{AAW_Perf2011} for a similar limiting analysis. Without loss of generality, we assume a unit mass of providers. For a provider at $[x,x+dx]$ with $x\in [0,1]$, we denote the capacity to be $C(x)$, the price to be $p(x)$ and the traffic served to be $q_c(x)$. We assume that $C(x) > 0$ for all $x\in [0,1]$ and $C=\int_0^1 C(x) dx$. Then the user equilibrium is given by
\begin{align}\label{eq:pceq}
p(x) + l_c\left(\frac{q_c(x)}{C(x)} \right) = P\left( \int_0^1 q_c(y) dy \right)=:\lambda \; \forall x \in [0,1] 
\end{align} 
The revenue of a provider at $[x,x+dx]$ is given by
\begin{align*}
R_c(x) = p(x) q_c(x) = q_c(x) \left( \lambda -  l_c\left(\frac{q_c(x)}{C(x)} \right)\right).
\end{align*}
Since the provider at $[x,x+dx]$ has no market power, we will assume that her price choice cannot change $\lambda$. For this scenario, we get the following result.
\begin{thm}
At the equilibrium between the providers, the user equilibrium is given by $\tfrac{q_c(x)}{C(x)} \equiv \bar{\alpha}$ (independent of $x\in[0,1]$) where $\bar{\alpha}$ is given by the unique solution in $[0, 1/C]$ of
\begin{align}
P(\alpha C) = l_c(\alpha) + \alpha  l_c^\prime(\alpha) 
\end{align}
The optimal price is independent of $x$ and is given by
\begin{align}
p_c(x) & = \frac{q_c(x)}{C(x)} l_c^\prime\left(\frac{q_c(x)}{C(x)} \right)\equiv\bar{\alpha}  l_c^\prime(\bar{\alpha}).
\end{align}
The marginal increase in total welfare with bandwidth allocation is given by
\begin{align}
\frac{\partial T_c}{\partial C} & = \bar{\alpha}^2 l_c^\prime(\bar{\alpha}).
\end{align}
Furthermore, the price chosen is such that the total welfare is also maximized, so that efficiency results.
\end{thm}
\begin{IEEEproof}
See Appendix~\ref{sec:competition}.
\end{IEEEproof}
The total revenue of all the providers, the total welfare and the consumer surplus are given by 
\begin{align*}
R_c & =\int_0^x R_c(x) dx = \bar{\alpha}^2  l_c^\prime(\bar{\alpha}) C, \ T  = \int_0^{\bar{\alpha} C} P(q) dq - \bar{\alpha} l_c(\bar{\alpha}) C, \\
S_c 
&  = \int_0^{\bar{\alpha} C} P(q) dq - \bar{\alpha} l_c(\bar{\alpha}) C - \bar{\alpha}^2  l_c^\prime(\bar{\alpha}) C. 
\end{align*}
When both the latency function and the demand function are linear, these simplify to
\begin{align}
\begin{split}
\bar{\alpha} & = \frac{1}{C + \frac{2}{p_{\max}}}, \ 
\bar{\alpha} C  = \frac{C p_{\max}}{C p_{\max}+2}, \ 
\frac{\partial T_c}{\partial C}  = \frac{1}{(C + \frac{2}{p_{\max}})^2},\\
R_c & = \frac{C}{\left(C+\frac{2}{p_{\max}}\right)^2}, \quad
S_c  = \frac{p_{\max}}{2} \left( \frac{C}{C+\frac{2}{p_{\max}}} \right)^2, \\
T_c & =\frac{C}{\left(C+\frac{2}{p_{\max}}\right)^2} +  \frac{p_{\max}}{2} \left( \frac{C}{C+\frac{2}{p_{\max}}} \right)^2.
\end{split}
\end{align}

We now generalize the perfect competition model by adding $W$ bandwidth as whitespace. 
\begin{prop}
At the equilibrium between the providers, the user equilibrium is given by $\tfrac{q_c(x)}{C(x)}\equiv\alpha_w$ and 
\begin{align*}
q_w^* = Q\big(\alpha_w l_c^\prime(\alpha_w) + l_c(\alpha_w)\big)-\alpha_w C,
\end{align*}
where $\alpha_w$ is the unique solution in $[0, \bar{\alpha}]$ of 
\begin{align*}
l_w\left( \frac{Q\big(\alpha l_c^\prime(\alpha) + l_c(\alpha)\big)-\alpha C}{W}\right) = \alpha l_c^\prime(\alpha) + l_c(\alpha).
\end{align*}
The optimal price is given by $p(x)\equiv \alpha_w l_c^\prime(\alpha_w)$.
\end{prop}
\begin{IEEEproof}
See Appendix~\ref{sec:competitionplus}.
\end{IEEEproof}
The total revenue of the providers, consumer surplus and total welfare are given by
\begin{align*}
R_{cw} & = \alpha_w^2 l_w^{\prime}(\alpha_w) C \\
S_{cw} & = \int_0^{\alpha_w C + q_w^* } P(q) dq - \big( \alpha_w l_c^\prime(\alpha_w) + l_c(\alpha_w)  \big) \big(\alpha_w C + q_w^*\big) \\
T_{cw} & = \int_0^{\alpha_w C + q_w^* } P(q) dq - \big( \alpha_w l_c^\prime(\alpha_w) + l_c(\alpha_w)  \big) q_w^* - l_c(\alpha_w) \alpha_w C 
\end{align*}
In the linear latencies and linear demand case, the expressions simplify to
\begin{align*}
\alpha_w & = \frac{1}{C+2W+\frac{2}{p_{\max}}}, \quad
\alpha_w C  = \frac{C}{C+2W+\frac{2}{p_{\max}}},\\
q_w^* & = \frac{2W}{C+2W+\frac{2}{p_{\max}}}, \quad \frac{\partial T_{cw}}{\partial W}  = \frac{8 W}{\left(C+2W+\frac{2}{p_{\max}} \right)^3}, \\
R_{cw} & = \frac{C}{\left(C+2W+\frac{2}{p_{\max}}\right)^2}, \quad
S_{cw}  = \frac{p_{\max}}{2} \left( \frac{C+2W}{C+2W+\frac{2}{p_{\max}}} \right)^2, \\
T_{cw} & =\frac{C}{\left(C+2W+\frac{2}{p_{\max}}\right)^2} +  \frac{p_{\max}}{2} \left( \frac{C+2W}{C+2W+\frac{2}{p_{\max}}} \right)^2, \\
\frac{\partial T_{cw}}{\partial C} & = \frac{C+6W+\frac{2}{p_{\max}}}{\left(C+2W+\frac{2}{p_{\max}} \right)^3}.
\end{align*}
Note that $\tfrac{\partial T_{cw}}{\partial W}\big|_{W=0} = 0$ in this case which contrasts with all other scenarios where a small amount of bandwidth is added as whitespace.

\section{Accounting For Investment Cost}\label{sec:invest}

In the analysis thus far, we have assumed that the extra spectrum is available to the monopolist free of cost. However, in reality, this spectrum would have to be purchased from another owner or from the government. We assume that the entity with spectrum resources is willing to assign $C_e$ units of bandwidth at a per unit price of $p_e$; the label $e$ is short-form for extra. We are mainly concerned with the case when the monopolist has an initial bandwidth endowment of $C>0$. However, we will also discuss the case when $C=0$. For simplicity we will analyze the all-linear setting, but in future work we will consider the more general setting in terms of demand and latency functions too. Since any purchase of spectrum counts against net revenue, we will use the characterization of optimal revenue based on the user equilibrium behavior characterized in Section~\ref{sec:results}.

The problem faced by the monopolist now is to maximize his net revenue, i.e., revenue obtained from the users minus the cost for the additional bandwidth. This is given by
\begin{align*}
\max_{c \in [0, C_e]} R_m(q_m^*(C+c)) - p_e c
\end{align*}
From our earlier expressions, this is a convex optimization problem. The derivative of the objective function is given by
\begin{align*}
\frac{p_{\max}}{4} \frac{1}{\left( (C+c) p_{\max} + 1\right)^2} - p_e.
\end{align*}
Therefore, the optimal purchase is given by
\begin{align*}
c^*_m = \min\left(\left[\frac{1}{2\sqrt{p_e}} - \frac{1}{p_{\max}} - C\right]_+,C_e\right).
\end{align*}
Using this, the maximum per unit price at which a monopolist will buy all the available spectrum, that is, the market clearing price, is given by
\begin{align*}
p_e^m = \frac{p_{\max}^2}{4 \left( (C+C_e) p_{\max} + 1 \right)^2}.
\end{align*}
We will use this for our comparisons. Note that this price can be used as a reference for the value of extra spectrum to the monopolist.

We now extend the analysis of this setting to consider the possibility of the monopolist buying extra spectrum. The monopolist once again maximizes her net revenue, but subject to her initial endowment of $C$ bandwidth units and with $W$ units of bandwidth used as whitespace. The optimization problem is given by
\begin{align*}
\max_{c \in [0, C_e]} R_m(q_w^*) - p_e c
\end{align*}
This is a simple extension of the previous analysis, and the optimal purchase is given by
\begin{align*}
c^*_w = \min\left(\left[\frac{1}{2\sqrt{p_e}} - \frac{1}{p_{\max}} - C-W\right]_+,C_e\right).
\end{align*}
Using this maximum per unit price at which a monopolist will buy all the available spectrum is given by
\begin{align*}
p_e^{mw} = \frac{p_{\max}^2}{4 \left( (C+C_e+W) p_{\max} + 1 \right)^2}.
\end{align*}
Note that for the purposes of buying new spectrum, the monopolist behaves as if the $W$ units of whitespace bandwidth are also part of her initial endowment.

Next we analyze the impact of investment cost of additional spectrum on competing service providers. Owing to competition, we will have a game formulation the equilibrium of which will yield the capacity investment choices. The timing of the game is as follows: the spectrum owner (say the government or some other such entity) releases up to $C_e^i$ units of bandwidth at unit price $p_e^i$ for $i=1,2$ to the two service providers; the service providers decide on the amount of spectrum to purchase $c_e^i$ for $i=1, 2$ with $C_i$ being their initial endowments; then they set prices and the users choose their access levels based upon the delivered price. Note that we allow for differential pricing. Building upon the results in Proposition~\ref{prop:Nc}, we seek a sub-game perfect equilibrium, and in particular, pure-strategy equilibria. 

We start by analyzing the revenue of the providers as a function of their capacity. 
\begin{lem}\label{lem:revenue}
The revenue of provider $i=1, 2$ at the Nash equilibrium characterized in Proposition~\ref{prop:Nc} is a strictly concave function of her capacity $C_i$.
\end{lem}
\begin{IEEEproof}
See Appendix~\ref{sec:concave}.
\end{IEEEproof}
As a consequence of Lemma~\ref{lem:revenue}, given the capacity of provider $-i$, the equilibrium prices, and the equilibrium user access levels, the best-response of provider $i$ is the (unique) solution of a convex optimization problem with a strictly concave objective function (the net revenue). Therefore, we obtain the following result characterizing the existence of a Nash equilibrium spectrum purchase strategy. 
\begin{thm}
The spectrum investment game is a concave game, so a pure strategy Nash equilibrium exists.
\end{thm}
\begin{IEEEproof}
The concave game property follows from Lemma~\ref{lem:revenue}, and then the result follows from standard existence theorems in \cite[Theorem 1]{FudenbergTiroleBook}, and \cite{Glicksberg1952, Rosen1965}.
\end{IEEEproof}

Since pure-strategy equilibria exist, the next quantities of interest are values of the prices $(p_e^1,p_e^2)$ such that at equilibrium the service providers buy the allocated capacities $(C_e^1, C_e^2)$. The social planner can then decide the best allocation of capacities (subject to available spectrum) to maximize the social welfare, which is the sum of the consumer surplus and the providers' revenues\footnote{Note that the investment cost term cancels in the social welfare calculation.}. From the best-response characterization, it follows that if the social planner chooses the prices to be $\bar{p}_e^i = \tfrac{\partial R_i}{\partial C_i}$ when the capacity allocation of the provider $i$ is given by $C_i + C_e^i$, then the equilibrium purchase of provider $i$ is, indeed, $C_e^i$. From Lemma~\ref{lem:revenue} we get that prices that the social planner should choose for $i=1, 2$ are given by
\begin{align*}
& \bar{p}_e^i = p_{\max}^3 \\
&\times \frac{\left(p_{\max} +  \frac{2}{C_{-i}+C_e^{-i}}\right)^2}{\left( \left(3 p^2_{\max} + \frac{4p_{\max}}{C_{-i}+C_e^{-i}}\right) \left(C_i+C_e^i\right) + 4 p_{\max} + \frac{4}{C_{-i}+C_e^{-i}}\right)} \\
& \times \frac{\left(7 p_{\max} + \frac{8}{C_{-i}+C_e^{-i}}\right) \left(C_i+C_e^i\right) +\frac{4 p_{\max}  + \frac{4}{C_{-i}+C_e^{-i}}}{3 p_{\max} + \frac{4}{C_{-i}+C_e^{-i}}} }{\left( \left(3 p^2_{\max} + \frac{4p_{\max}}{C_{-i}+C_e^{-i}}\right) \left(C_i+C_e^i\right) + 4 p_{\max} + \frac{4}{C_{-i}+C_e^{-i}}\right)^2}.
\end{align*}

Using the same analysis procedure as in the competitive setting, we can show that a pure-strategy equilibrium exists when there some bandwidth set-aside as whitespace, and we can also determine the unit price of spectrum at which the providers purchase the entirety of the allocated spectrum. In the interest of brevity, we omit the details. 

For the perfect competition case, we conjecture that any additional spectrum should be handed out free of charge, and for simplicity, proportional to the initial endowment. We believe that many other spectrum allocation mechanisms will yield the same social welfare. Proving this conjecture is for future work.

\section{Discussion}\label{sec:discuss}

The value of the results in Section~\ref{sec:results} is in providing comparative statics for the different market scenarios being considered for new spectrum. The contribution of Section~\ref{sec:invest} is in proposing appropriate per unit valuations for any additional spectrum based on the different market scenarios. For all our numerical results, we will always assume that the latency function is linear and same for every player (provider or whitespace). We will also assume a linear demand function on the user side with with $q_{\max}=1$ and $p_{\max}=1$.

\begin{figure}[htbp] 
   \centering
   \includegraphics[width=3.25in,height=2in]{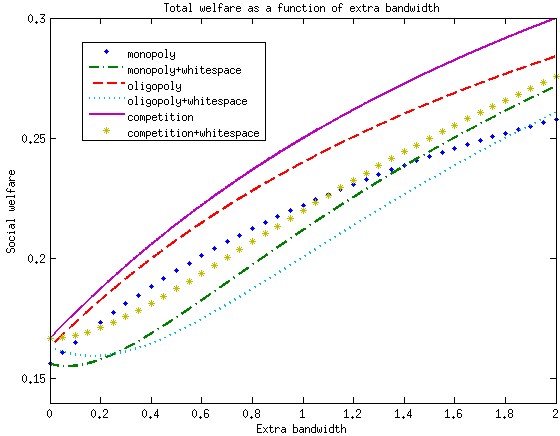} 
   \caption{Total welfare as a function of bandwidth allocated according to the different schemes for a particular choice of linear latencies and demand function}
   \label{fig:specalloc}
\end{figure}

For the first set of results, we will assume that we have initial allocated bandwidth of $1$ unit. Here we explore the total welfare achieved under the different market types when an additional bandwidth of up to $2$ units is allocated. The total welfare as a function of the additional bandwidth is shown in Figure~\ref{fig:specalloc}. The different curves correspond to the results in Section~\ref{sec:results} and cover the cases of monopoly provider, monopoly provider with extra bandwidth only as open access, oligopolistic providers with with an initial endowment of $1/2$ units of capacity and the extra capacity allocated equally, oligopolistic providers with an initial endowment of $1/2$ units of capacity competing with extra bandwidth only as open access, perfectly competitive market (with $C(x)\equiv 1, \forall x\in [0,1]$) and perfectly competitive market ($\forall x\in [0,1],C(x)\equiv 1$) with additional bandwidth only as open-access. For these settings, we will be using the expressions obtained for linear latency and demand functions in Section~\ref{sec:results}.

Some salient features of the curves in Figure~\ref{fig:specalloc} are the following. Adding a small amount of additional bandwidth only as open-access results in a decrease in total welfare for both the monopolistic and oligopolistic settings, and here it would be better to allocate the bandwidth to the service providers. With a sufficiently large bandwidth allocation, the total welfare is actually better when allocating the spectrum as unlicensed access when there is a monopoly provider. In the perfectly competitive setting, adding additional bandwidth only as open-access leads to a much slower increase in total welfare, nevertheless, the total welfare does increase. Within the settings of our model, it is better to allocate all the bandwidth to the service providers instead of setting it aside as whitespace, even in the oligopolistic setting. A final point is that having even two competing providers brings the outcome much closer to the efficient outcome seen in the perfectly competitive situation.

\begin{figure}[htbp] 
   \centering
   \includegraphics[width=3.25in,height=1.5in]{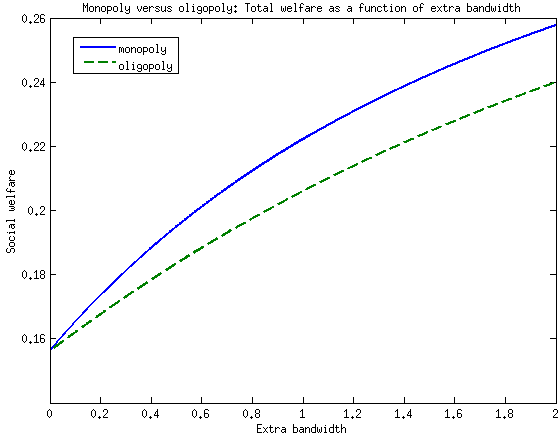} 
   \caption{Total welfare as a function of bandwidth allocated according to the different schemes for a particular choice of linear latencies and demand function}
   \label{fig:comp}
\end{figure}

Next we consider case where there is a monopolist with an initial capacity endowment of $C=1$. From Figure~\ref{fig:specalloc}, we know that allocating a small amount of bandwidth as whitespace reduces the total welfare and only when a significantly large amount of bandwidth is allocated as whitespace, is the total welfare better than allocating it to the monopolist. However, this bandwidth can be allocated to a new service provider who will compete with the incumbent. In Figure~\ref{fig:comp}, we compare the total welfare obtained by allocating all the new bandwidth to a new competitor with allocating it to the incumbent. Surprisingly, we find that in terms of total welfare, it is better to allocate the new bandwidth to the monopolist that to create a new competing service provider. 

\begin{figure}[htbp] 
   \centering
   \includegraphics[width=3.25in,height=1.5in]{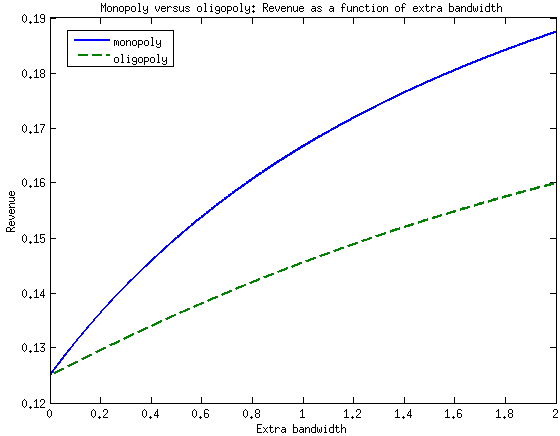} 
   \caption{Total welfare as a function of bandwidth allocated according to the different schemes for a particular choice of linear latencies and demand function}
   \label{fig:compr}
\end{figure}

\begin{figure}[htbp] 
   \centering
   \includegraphics[width=3.25in,height=1.5in]{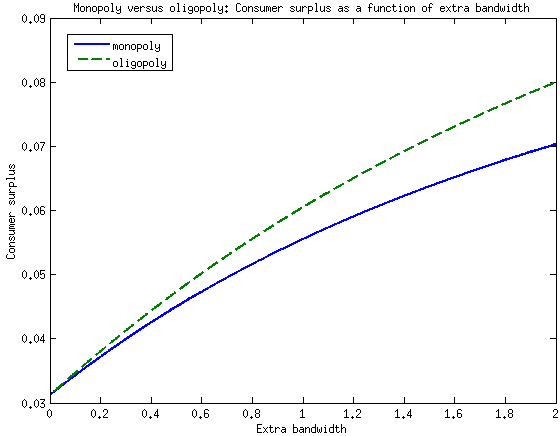} 
   \caption{Total welfare as a function of bandwidth allocated according to the different schemes for a particular choice of linear latencies and demand function}
   \label{fig:comps}
\end{figure}

In Figures~\ref{fig:compr} and \ref{fig:comps}, we look at this case in greater detail by plotting the revenue and the consumer surplus, respectively. From the figures it follows that creating a new competing service provider leads to much higher consumer surplus, but the increase in total revenue of the service provider(s) is much than when the bandwidth is allocated to the monopolist. Thus, if consumer surplus is what drives policy, then it is better to create the new competing service provider.

\begin{figure}[htbp] 
   \centering
   \includegraphics[width=3.25in,height=1.5in]{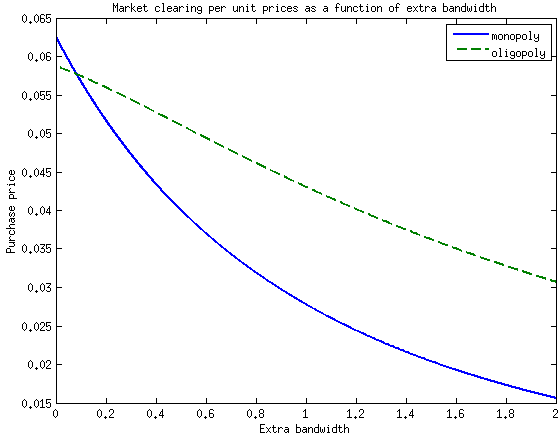} 
   \caption{Total welfare as a function of bandwidth allocated according to the different schemes for a particular choice of linear latencies and demand function}
   \label{fig:price}
\end{figure}

Finally, we compare the different market scenarios in terms of the market clearing price when the providers have to purchase spectrum from another entity (government or private). In Figure~\ref{fig:price} we plot the market clearing per unit price from Section~\ref{sec:invest} for two cases when up to two units of bandwidth can be allocated: first, the monopoly setting where the monopolist has an initial endowment of $C=1$; and second, the oligopoly setting where the competing providers have equal initial endowments of $C_1=C_2=1/2$ units and any initial spectrum is allocated equally. If the amount of spectrum to allocated is small, then in terms of revenue of the sale of spectrum, the monopoly setting is better. However, if maximizing total welfare is the goal, then from the plots in Figure~\ref{fig:specalloc}, the oligopolistic setting is better. Thus, it follows that a social planner (or entity like the government) would always prefer the oligopoly setting, but a private seller with a small amount to spectrum to sell would prefer the monopoly setting.

The model we've considered ignores many important factors that could dramatically change the conclusions. First, we assume that all bandwidth that has been allocated and is available for allocation is equally good. Given the different propagation characteristics for different frequencies, such a simplistic assumption needs to be revisited by allowing for heterogeneity in latency functions depending on the frequency. The heterogeneity of the frequency bands also impacts the provisioning costs (network architecture, number of access points, number of repeaters, cost of devices, etc.), and so a more comprehensive model would reflect these differences. Considering heterogeneity of the providers in terms of latency functions and market power would also be important. 
Additionally, the extraction of utility from a particular frequency band also depends on the applications using the band. The 2.4GHz and 5GHz bands were not considered commercially viable before the advent and success of 802.11/WiFi~\cite{EilatLevinMilgrom2011,Benkler2012}. A coupling of innovation cycles with the bandwidth provisioning methods chosen would be desirable to model. Another related aspect that we do not model in this analysis is the increase of the customer base with increasing bandwidth allocation, although by assuming a large enough $q_{\max}$ this effect can easily be incorporated. For the sake of simplicity and to facilitate explicit calculations, we restricted the number of competing providers to two, and considering more than two would be valuable as well. We have also assumed a homogeneous user population. At the very least, it would be useful to model a heterogeneous population such as in \cite{NZBHV_N2011}. Finally, we have not modeled any uncertainty in our analysis, in terms of knowledge of the population base, demand functions, or bandwidth availability. In future work, we plan to address these shortcomings with a more detailed and realistic model. 

\section{Conclusions}\label{sec:conc}

Using the model of competition in congestible resources, we analyzed the impact of adding additional spectrum under different market conditions. Specifically, we contrasted monopolistic, oligopolistic and perfectly competitive spectrum markets with and without additional unlicensed or whitespace spectrum. Using the equilibrium analysis of these scenarios, we then considered a simple model of how the providers would consider additional investment in spectrum. We then used this model to propose appropriate per unit valuations in each scenario.

\appendices

\section{Monopoly Case}\label{sec:monopoly}

For a given price $p \geq 0$, the amount of traffic that arrives $q_m$ is given by the solution of 
\begin{align*}
p + l_m(q_m/C) = P(q_m)
\end{align*}
if $p \in [0, p_{\max}]$, and otherwise $q_m\equiv0$. For revenue maximization it is clear that $p \in [0, p_{\max}]$ so we can assume that $p= P(q_m) - l_m(q_m/C)$ which is a concave decreasing function of $q_m$; for later use denote $p_m(q_m) :=P(q_m) - l_m(q_m/C)$. Note that $q_m$ is greatest when $p=0$ and this value is given by $\hat{q}_m$ the unique non-negative solution to $l_m(q/C) = P(q)$. The revenue maximization problem is then
\begin{align*}
\max_{p_m \in [0, p_{\max}]} R_m(p) = q_m p_m, 
\end{align*}
which can be rewritten as
\begin{align*}
\max_{q_m \in [0, \hat{q}_m]} R_m(q_m)=q_m ( P(q_m) - l_m(q_m/C) )
\end{align*}
It is easily verified that the revenue $R_m(q_m)$ is a strictly concave function of $q_m$.
Thus, there is a unique maximizer which can also be shown to be an interior solution. Let the maximizer be $q_m^*$, then the revenue optimal asking price $p_m^*=P(q_m^*)-l_c(q_m^*/C)$. Furthermore, $q_m^*$ satisfies
\begin{align*}
q_m^*= - \frac{P(q_m^*)-l_m\left(\frac{q_m^*}{C}\right)}{P^\prime(q_m^*) - \frac{l_m^\prime\left(\frac{q_m^*}{C}\right)}{C}}
\end{align*}
This relationship can now be used to determine the change in quantity served with change in $C$. Working through the details we get
\begin{align*}
& \frac{\partial q_m^*}{\partial C} =\\
& \frac{\left(P(q_m^*) - l_m\left(\frac{q_m^*}{C}\right)\right) \left( 2 \frac{ l_m^\prime\left(\frac{q_m^*}{C}\right) }{C^2} + \frac{q_m^*l_m^{\prime\prime}\left(\frac{q_m^*}{C}\right) }{C^3} \right)  }{\beta},\\
& \text{where } \beta:=2 \left(P^\prime(q_m^*)-\frac{l_m^\prime\left(\frac{q_m^*}{C}\right)}{C}\right)^2 \\
& \qquad \quad - \left(P(q_m^*) - l_m\left(\frac{q_m^*}{C}\right)\right) \left( P^{\prime\prime}(q_m^*) - \frac{l_m^{\prime\prime}\left(\frac{q_m^*}{C} \right)}{C^2}\right).
\end{align*}
It is easy to see that $\tfrac{\partial q_m^*}{\partial C} > 0$. Using the characterization of $p_m^*$ we have
\begin{align*}
& \frac{\partial p_m^*(q_m^*)}{\partial C}  = \frac{\partial q_m^*}{\partial C} \left( P^\prime(q_m^*) - \frac{l_m^\prime\left(\frac{q_m^*}{C}\right)}{C}\right) + q_m^* \frac{l_m^\prime\left(\frac{q_m^*}{C}\right)}{C^2} \\
& = \frac{\left(P(q_m^*) - l_m\left(\frac{q_m^*}{C}\right)\right) q_m^* }{\beta} \\
& \quad \times \left( P^\prime(q_m^*)\frac{l_m^{\prime\prime}\left(\frac{q_m^*}{C}\right) }{C^3}  -P^{\prime\prime}(q_m^*)  \frac{l_m^\prime\left(\frac{q_m^*}{C}\right)}{C^2}  \right).
\end{align*}
The sign of $\tfrac{\partial p_m(q_m^*)}{\partial C}$ depends on particular functions involved and the value of $C$. However, we note that for linear price and latency functions, the monopoly price does not depend on $C$ as the derivative is $0$!

The consumer surplus for a given price $p\in [0,p_{\max}]$ and corresponding quantity $q_m$ is given by
\begin{align*}
S_m(q_m) 
= \int_0^{q_m} P(q) dq - q_m P(q_m)
\end{align*}
Therefore, the total welfare is given by 
\begin{align*}
T_m(q_m) = S_m(q_m) + R_m(q_m) = \int_0^{q_m} P(q) dq - q_m l_m\left(\frac{q_m}{C} \right)
\end{align*}
At the revenue optimal operating point $q_m^*$, these are given by $R_m(q_m^*)$, $S_m(q_m^*)$ and $T(q_m^*)$ respectively. The change in total welfare is given by
\begin{align*}
& \frac{\partial T_m(q_m)}{\partial C} \\
&   = \frac{\partial q_m^*}{\partial C} \left( P(q_m^*) - l_m\left( \frac{q_m^*}{C}\right) - q_m^* \frac{l_m^\prime\left(\frac{q_m^*}{C}\right)}{C}\right) + q_m^* \frac{l_m^\prime\left(\frac{q_m^*}{C}\right)}{C^2} \\
& = - q_m^* P^\prime(q_m^*) \frac{\partial q_m^*}{\partial C} + q_m^* \frac{l_m^\prime\left(\frac{q_m^*}{C}\right)}{C^2}  
\end{align*}
which is positive. Thus, both the quantity served and total welfare increase when the capacity $C$ is increased.

\section{Monopoly with incremental white-space}\label{sec:monopolyplus}

Here we consider the addition of an infinitesimal addition of whitespace when there is a monopolist. Let $\delta W$ be this amount of whitespace. Then the user equilibrium condition implies
\begin{align*}
p+l_m\left( \frac{q_m}{C}\right) = l_w\left( \frac{q_w}{\delta W}\right) = P(q_m+q_w)
\end{align*}
Since we're adding only a small amount of whitespace and since all the functions used are continuously differentiable, we will always be comparing with the monopoly setting. Thus, we can define the following
\begin{align*}
p  = p_m^* + \delta p, \quad
q_m  = q_m^* + \delta q_m, \quad
\frac{q_w}{\delta W}  = \hat{q}_w + \delta q_w
\end{align*}
The reason for the different definition of $q_w$ will become clear very soon. Feeding this into the user equilibrium condition we (approximately) get
\begin{align*}
& p_m^*+l_c\left( \frac{q_m^*}{C}\right) + \delta p + \delta q_m \frac{l_m^\prime\left( \frac{q_m^*}{C}\right)}{C} + \delta^2 q_m \frac{l_m^{\prime\prime}\left( \frac{q_m^*}{2C^2}\right)}{C} \\
& = l_w(\hat{q}_w) + \delta q_w l_w^\prime(\hat{q}_w) + \delta^2 q_w  \frac{l_w^{\prime\prime}(\hat{q}_w)}{2} \\
& = P(q^*_m) + (\delta q_m + \hat{q}_w \delta W) P^\prime(q^*_m) + (\delta q_m + \hat{q}_w \delta W)^2 \frac{P^{\prime\prime}(q^*_m)}{2} \\
& = P(q^*_m) + (\delta q_m + \hat{q}_w \delta W) P^\prime(q^*_m) \\
& \qquad + ( \delta^2 q_m + 2 \hat{q}_w \delta W \delta q_m + \hat{q}^2_w \delta^2 W) \frac{P^{\prime\prime}(q^*_m)}{2}
\end{align*}
Equating the non-infinitesimal terms and the infinitesimal terms we get
\begin{align*}
& \hat{q}_w   = l_w^{-1}\left(p_m^*+l_m\left( \frac{q_m^*}{C}\right) \right) = l_w^{-1}(P(q_m^*)) \\
 & \delta p   = \delta q_m \left( P^\prime(q^*_m)  - \frac{l_m^\prime\left( \frac{q_m^*}{C}\right)}{C} \right) + \hat{q}_w \delta W P^\prime(q^*_m) \\
& \quad + \delta q_w \delta W P^\prime(q^*_m)  + \left( \delta^2 q_m + 2 \hat{q}_w \delta W \delta q_m + \hat{q}^2_w \delta^2W\right) \frac{P^{\prime\prime}(q^*_m)}{2}\\
& \delta q_w l_w^\prime(\hat{q}_w) + \delta^2 q_w  \frac{l_w^{\prime\prime}(\hat{q}_w)}{2}    = \delta p + \delta q_m \frac{l_m^\prime\left( \frac{q_m^*}{C}\right)}{C} + \delta^2 q_m \frac{l_m^{\prime\prime}\left( \frac{q_m^*}{2C^2}\right)}{C}
\end{align*}
For analytical tractability we will start by assuming that $l_w(\cdot)$ is a linear function so that its second derivative is $0$, later on we will show that the general case can be analyzed in exactly the same manner. Using this assumption we then get
\begin{align*}
\delta q_w &  = \frac{\delta p + \delta q_m \frac{l_m^\prime\left( \frac{q_m^*}{C}\right)}{C} + \delta^2 q_m \frac{l_m^{\prime\prime}\left( \frac{q_m^*}{2C^2}\right)}{C}}{l_w^\prime(\hat{q}_w)} \\
&  \approx \frac{\delta p + \delta q_m \frac{l_m^\prime\left( \frac{q_m^*}{C}\right)}{C} }{l_w^\prime(\hat{q}_w)},
\end{align*}
where we ignored the higher order term since $\delta q_w$ gets multiplied by $\delta W$.

Using the last equation to substituting for $\delta q_w$ in the expression for $\delta p$ and ignoring higher order terms, we get
\begin{align*}
& \delta p  \approx \frac{\delta q_m \left( P^\prime(q^*_m)  - \frac{l_m^\prime\left( \frac{q_m^*}{C}\right)}{C} \right)  }{1-\delta W\frac{P^\prime(q_m^*)}{l_w^\prime(\hat{q}_w)} }\\
&  + \frac{ \hat{q}_w \delta W P^\prime(q^*_m) + \delta q_m \delta W\frac{l_m^\prime\left( \frac{q_m^*}{C}\right)}{C}\frac{P^\prime(q_m^*)}{l_w^\prime(\hat{q}_w)} }{1-\delta W\frac{P^\prime(q_m^*)}{l_w^\prime(\hat{q}_w)} }\\
&  +  \frac{( \delta^2 q_m + 2 \hat{q}_w \delta W \delta q_m + \hat{q}^2_w \delta^2W) \frac{P^{\prime\prime}(q^*_m)}{2}}{1-\delta W\frac{P^\prime(q_m^*)}{l_w^\prime(\hat{q}_w)} } \\
& \approx \delta q_m \left( P^\prime(q^*_m)  - \frac{l_m^\prime\left( \frac{q_m^*}{C}\right)}{C} \right) + \hat{q}_w \delta W P^\prime(q^*_m) \\
&  + \hat{q}_w \delta^2W \left( P^\prime(q^*_m) \frac{P^\prime(q_m^*)}{l_w^\prime(\hat{q}_w)} + \frac{P^{\prime\prime}(q^*_m)}{2}\right) + \delta q_m \delta W \hat{q}_w P^{\prime\prime}(q^*_m)\\
& + \delta q_m \delta W \frac{P^\prime(q_m^*)}{l_w^\prime(\hat{q}_w)} \left( P^\prime(q^*_m)  - \frac{l_m^\prime\left( \frac{q_m^*}{C}\right)}{C} \right)  \\
&  + \delta q_m \delta W \frac{l_m^\prime\left( \frac{q_m^*}{C}\right)}{C}\frac{P^\prime(q_m^*)}{l_w^\prime(\hat{q}_w)}+ \delta^2 q_m \frac{P^{\prime\prime}(q^*_m)}{2}\\
& = \delta q_m \left( P^\prime(q^*_m)  - \frac{l_m^\prime\left( \frac{q_m^*}{C}\right)}{C} \right) + \hat{q}_w \delta W P^\prime(q^*_m) \\
&  + \hat{q}_w \delta^2W \left( P^\prime(q^*_m) \frac{P^\prime(q_m^*)}{l_w^\prime(\hat{q}_w)} + \frac{P^{\prime\prime}(q^*_m)}{2}\right) \\
&  + \delta q_m \delta W \left(\frac{P^\prime(q_m^*)}{l_w^\prime(\hat{q}_w)} P^\prime(q^*_m)+ \hat{q}_w P^{\prime\prime}(q^*_m)\right) + \delta^2 q_m \frac{P^{\prime\prime}(q^*_m)}{2}
\end{align*}

The revenue is now given by
\begin{align*}
& R_{mw}  = p q_m = (p_m^* + \delta p) (q_m^* + \delta q_m) \\
& = p_m^*q_m^* + \delta q_m p_m^* + \delta p q_m^* + \delta p \delta q_m \\
& = p_m^*q_m^* + \delta q_m \left(p_m^* + q_m^*\left( P^\prime(q^*_m)  - \frac{l_m^\prime\left( \frac{q_m^*}{C}\right)}{C} \right)  \right) \\
&  + q_m^* \hat{q}_w \delta W P^\prime(q^*_m) + \hat{q}_w q_m^* \delta^2W \left( P^\prime(q^*_m) \frac{P^\prime(q_m^*)}{l_w^\prime(\hat{q}_w)} + \frac{P^{\prime\prime}(q^*_m)}{2}\right)\\
&  + \delta^2q_m \left( P^\prime(q^*_m)  - \frac{l_m^\prime\left( \frac{q_m^*}{C}\right)}{C}+  q_m^* \frac{P^{\prime\prime}(q^*_m)}{2}\right) \\
&  + \delta q_m \delta W \left( P^\prime(q^*_m) \left(\hat{q}_w +  q_m^*\frac{P^\prime(q_m^*)}{l_w^\prime(\hat{q}_w)}\right) + q_m^* \hat{q}_w P^{\prime\prime}(q^*_m)\right),
\end{align*}
where we ignore higher order terms.
Since $p_m^*=P(q_m^*) - l_m\left( \tfrac{q_m^*}{C}\right)$, using the characterization of $q_m^*$ we can simplify the expression above to
\begin{align*}
& R_{mw}  = R^*_m + q_m^* \hat{q}_w \delta W P^\prime(q^*_m)\\
&  +  \hat{q}_w q_m^* \delta^2W \left( P^\prime(q^*_m) \frac{P^\prime(q_m^*)}{l_w^\prime(\hat{q}_w)} + \frac{P^{\prime\prime}(q^*_m)}{2}\right)\\
&  + \delta^2q_c \left( P^\prime(q^*_m)  - \frac{l_c^\prime\left( \frac{q_m^*}{C}\right)}{C}+  q_m^* \frac{P^{\prime\prime}(q^*_m)}{2}\right) \\
& + \delta q_c \delta W \left( P^\prime(q^*_m) \left(\hat{q}_w +  q_m^*\frac{P^\prime(q_m^*)}{l_w^\prime(\hat{q}_w)}\right) + q_m^* \hat{q}_w P^{\prime\prime}(q^*_m)\right)
\end{align*}
Using the properties of $P(\cdot)$ and $l_m(\cdot)$, it is clear that the revenue is a concave function of $\delta q_m$; without the concavity assumption on $P(\cdot)$ we need $P^\prime(q^*_m)  -l_m^\prime\left( \tfrac{q_m^*}{C}\right)/C+  q_m^* \tfrac{P^{\prime\prime}(q^*_m)}{2} < 0$ for concavity in $\delta q_m$. Setting the derivative to be zero yields
\begin{align*}
\delta q_m^* = -\frac{1}{2}\frac{\left(\hat{q}_w +  q_m^*\frac{P^\prime(q_m^*)}{l_w^\prime(\hat{q}_w)}\right) P^\prime(q^*_m) + q_m^* \hat{q}_w P^{\prime\prime}(q^*_m)}{P^\prime(q^*_m)  - \frac{l_m^\prime\left( \frac{q_m^*}{C}\right)}{C}+  q_m^* \frac{P^{\prime\prime}(q^*_m)}{2}} \delta W
\end{align*}
It is easy to see that $\delta q_m^*$ is negative if $\hat{q}_w +  q_m^*\tfrac{P^\prime(q_m^*)}{l_w^\prime(\hat{q}_w)}\geq 0$, so that adding a small amount of whitespace results in reduction in the traffic served by the monopolist when the revenue maximization procedure is carried out. Having identified $\delta q_m^*$, the infinitesimal change in revenue after ignoring higher order terms is
\begin{align*}
\delta R_{mw} = q_m^* \hat{q}_w \delta W P^\prime(q^*_m). 
\end{align*}
It is easy to see that the revenue of the provider decreases with the addition of whitespace.

The consumer surplus is given by
\begin{align*}
& S_{mw} = \int_{0}^{q_c+q_w} P(q) dq - (q_c+q_w) P(q_c+q_w) \\
& = \int_0^{q_m^*} P(q) dq - q_m^* P(q_m^*) + (\delta q_c + \delta W \hat{q}_w) P(q_m^*) \\
&  - (\delta q_c + \delta W \hat{q}_w) (P(q_m^*)+q_m^* P^\prime(q_m^*))
\end{align*}
so that the infinitesimal change in consumer surplus at the revenue maximization point after ignoring higher order terms is
\begin{align*}
& \delta S_{mw}  =- (\delta q_c^* + \delta W \hat{q}_w) q_m^* P^\prime(q_m^*) \\
& = -\delta W q_m^* P^\prime(q_m^*)\\
&  \times \left(-\frac{1}{2}\frac{\left(\hat{q}_w +  q_m^*\frac{P^\prime(q_m^*)}{l_w^\prime(\hat{q}_w)}\right) P^\prime(q^*_m) + q_m^* \hat{q}_w P^{\prime\prime}(q^*_m)}{P^\prime(q^*_m)  - \frac{l_m^\prime\left( \frac{q_m^*}{C}\right)}{C}+  q_m^* \frac{P^{\prime\prime}(q^*_m)}{2}} + \hat{q}_w \right)  \\
\end{align*}

Using the two quantities derived above, we get the infinitesimal change in total welfare to be
\begin{align*}
& \delta T_{mw}  = \delta R + \delta S \\
& = \delta W \frac{1}{2} \frac{\left(\hat{q}_w +  q_m^*\frac{P^\prime(q_m^*)}{l_w^\prime(\hat{q}_w)}\right) P^\prime(q^*_m) + q_m^* \hat{q}_w P^{\prime\prime}(q^*_m)}{P^\prime(q^*_m)  - \frac{l_c^\prime\left( \frac{q_m^*}{C}\right)}{C}+  q_m^* \frac{P^{\prime\prime}(q^*_m)}{2}} q_m^* P^\prime(q_m^*)
\end{align*}
Again, it is easy to verify that $\delta T_{mw}$ is negative if $\hat{q}_w +  q_m^*\tfrac{P^\prime(q_m^*)}{l_w^\prime(\hat{q}_w)}\geq 0$. Finally, we get 
\begin{align*}
& \frac{\partial T_{mw}}{\partial W}\Big|_{W=0}  =  \\
& \frac{1}{2} \frac{\left(\hat{q}_w +  q_m^*\frac{P^\prime(q_m^*)}{l_w^\prime(\hat{q}_w)}\right) P^\prime(q^*_m) + q_m^* \hat{q}_w P^{\prime\prime}(q^*_m)}{P^\prime(q^*_m)  - \frac{l_m^\prime\left( \frac{q_m^*}{C}\right)}{C}+  q_m^* \frac{P^{\prime\prime}(q^*_m)}{2}} q_m^* P^\prime(q_m^*)
\end{align*}
Note that our analysis also yields the partial derivatives of $p$, $q_c$, $R_{mw}$ and $S_{mw}$ at $W=0$.


Further restricting the latency function $l_m(\cdot)$ to be linear, i.e., $l_m(x)=x$, we have using the characterization of the monopoly setting
\begin{align*}
q_m^* P^\prime(q_m^*) + P(q_m^*)  = 2 \frac{q_m^*}{C}, \quad
\hat{q}_w  = P(q_m^*).
\end{align*}
Using these we have
\begin{align*}
& \hat{q}_w +  q_m^*\frac{P^\prime(q_m^*)}{l_w^\prime(\hat{q}_w)}  = P(q_m^*)+ q_m^* P^\prime(q_m^*)  = 2 \frac{q_m^*}{C} \\
& \frac{\partial T_{mw}}{\partial W}\Big|_{W=0} =   \frac{1}{2} \frac{ 2 \frac{P^\prime(q^*_m)}{C} + P(q_m^*) P^{\prime\prime}(q^*_m)}{P^\prime(q^*_m)  - \frac{l_m^\prime\left( \frac{q_m^*}{C}\right)}{C}+  q_m^* \frac{P^{\prime\prime}(q^*_m)}{2}} \Big(q_m^*\Big)^2 P^\prime(q_m^*)
\end{align*}
Now it is easy to see that $\tfrac{\partial T_{mw}}{\partial W}\Big|_{W=0}$ is negative.

In the general nonlinear whitespace latency case, $\delta q_w$ is obtained by solving the following quadratic equation,
\begin{align*}
 \delta q_w l_w^\prime(\hat{q}_w) + \delta^2 q_w  \frac{l_w^{\prime\prime}(\hat{q}_w)}{2}  = \delta p + \delta q_c \frac{l_c^\prime\left( \frac{q_m^*}{C}\right)}{C} + \delta^2 q_c \frac{l_c^{\prime\prime}\left( \frac{q_m^*}{2C^2}\right)}{C}
\end{align*}
The only viable solution is given by
\begin{align*}
& \delta q_w  = \frac{ - l_w^\prime(\hat{q}_w) }{l_w^{\prime\prime}(\hat{q}_w)} + \\
& \frac{\sqrt{\left(l_w^\prime(\hat{q}_w)\right)^2+2 l_w^{\prime\prime}(\hat{q}_w) \left( \delta p + \delta q_m \frac{l_m^\prime\left( \frac{q_m^*}{C}\right)}{C} + \delta^2 q_m \frac{l_m^{\prime\prime}\left( \frac{q_m^*}{2C^2}\right)}{C} \right) } }{l_w^{\prime\prime}(\hat{q}_w)} \\
& \approx \frac{\delta p + \delta q_m \frac{l_c^\prime\left( \frac{q_m^*}{C}\right)}{C} + \delta^2 q_m \frac{l_m^{\prime\prime}\left( \frac{q_m^*}{2C^2}\right)}{C}}{l_w^\prime(\hat{q}_w)} \\
& - \frac{l_w^{\prime\prime}(\hat{q}_w) \left( \delta p + \delta q_m \frac{l_m^\prime\left( \frac{q_m^*}{C}\right)}{C} + \delta^2 q_m \frac{l_m^{\prime\prime}\left( \frac{q_m^*}{2C^2}\right)}{C} \right)^2}{\left(l_w^\prime(\hat{q}_w)\right)^3}
\end{align*}
Since $\delta q_w$ gets multiplied by $\delta W$ and since we don't need to account for higher than second order terms, it suffices to approximate $\delta q_w$ as
\begin{align*}
\delta q_w \approx \frac{\delta p + \delta q_m \frac{l_m^\prime\left( \frac{q_m^*}{C}\right)}{C} }{l_w^\prime(\hat{q}_w)},
\end{align*}
which is exactly the form that we derived when $l_w(\cdot)$ was assumed to be linear so that the same result holds.


\section{Competing providers}\label{sec:oligopoly}

Let the capacities of the two providers be $C_1$ and $C_2$, respectively. We assume that they charge prices $p_1$ and $p_2$ and get quantities $q_1$ and $q_2$, respectively. It is easy to argue that neither provider will charge a price so high that all the customers go with the other. Thus, we can restrict attention to the case where $q_1, q_2 > 0$. In such scenario, the user equilibrium is given by
\begin{align*}
p_1 + \frac{q_1}{C_1} = p_2 + \frac{q_2}{C_2} = P(q_1+q_2)
\end{align*}
In the analysis we will ignore some of the edge cases but justify from the final solution as to why they are not necessary to consider. For $i\in\{1,2\}$, denote be $-i$ the set $\{1,2\}\setminus \{i\}$. From the user equilibrium, we find that the prices for $i=1, 2$ are given by
\begin{align*}
p_i & = p_{\max} (1-q_{-i}) - q_i\left(p_{\max} +\frac{1}{C_i} \right). 
\end{align*}
Given $(q_1,q_2)$, the prices are determined so we analyze the strategy of the providers by considering $(q_1,q_2)$ instead; ensuring non-negativity of prices and other such considerations are what we will ignore for the moment. The revenue of the providers for $i=1, 2$ is then given by
\begin{align*}
R_i(q_i,q_{-i}) & = p_{\max} (1-q_{-i}) q_i - q^2_i\left(p_{\max} +\frac{1}{C_i} \right)
\end{align*} 
Note that provider $i\in\{1,2\}$ maximizes her revenue by choosing $q_i$ given the value of $q_{-i}$. The best response of each provider $i=1, 2$ is then given by
\begin{align*}
\hat{q}_i(q_{-i}) & = \frac{p_{\max}}{2} \frac{1-q_{-i}}{p_{\max} +\frac{1}{C_i}}
\end{align*}
The unique fixed point of the above two equations for $i=1, 2$ is given by
\begin{align*}
q_i^* & = \frac{p_{\max}^2 + 2 \frac{p_{\max}}{C_{-i}}}{3 p_{\max}^2 + 4 p_{\max} \left(\frac{1}{C_1}+\frac{1}{C_2} \right) + \frac{4}{C_1 C_2}} 
\end{align*}
The corresponding prices for $i=1, 2$ are
\begin{align*}
p_i^* & = p_{\max} \frac{p_{\max}^2 + \frac{p_{\max}}{C_i}+2 \frac{p_{\max}}{C_{-i}}  + \frac{2}{C_1 C_2}}{3 p_{\max}^2 + 4 p_{\max} \left(\frac{1}{C_1}+\frac{1}{C_2} \right) + \frac{4}{C_1 C_2}}
\end{align*}
It is easy to verify that $(p_1^*,p_2^*)$ is a Nash equilibrium. Additionally note that $p_i^* < p_{\max}/2 = p_m^*$ for $i=1,2$, i.e., the asking price is strictly lower than the monopoly price which is a consequence of competition.

As mentioned earlier, the figures of merit are now given by
\begin{align*}
R_1(p_1^*,p_2^*) & = p_1^* q_1^*, \quad R_2(p_2^*,p_1^*)  = p_2^* q_2^*, \\
S_o(p_1^*,p_2^*) & = \frac{p_{\max}}{2} (q_1^*+q_2^*)^2, \\
T_o(p_1^*,p_2^*) & = R_1(p_1^*,p_2^*) + R_2(p_2^*,p_1^*) + S(p_1^*,p_2^*)
\end{align*}
We do not give the expressions as they're quite unwieldy.

\section{Competing providers with white-space}\label{sec:oligopolyplus}

Now we consider the addition of $W$ units of whitespace. As before, let $q_w$ be the quantity served in the whitespace. Again, at the user equilibrium (assuming non-zero quantities) we have
\begin{align*}
p_1 + \frac{q_1}{C_1} = p_2 + \frac{q_2}{C_2} = \frac{q_w}{W} = P(q_1+q_2+q_w)
\end{align*}
Using these equations we have
\begin{align*}
\forall i=1, 2 \quad p_i & = p_{\max} (1-q_w-q_{-i}) - q_i\left(p_{\max} +\frac{1}{C_i} \right), \\
q_w & = \frac{1-q_1-q_2}{1+\frac{1}{W p_{\max}}}
\end{align*}
Substituting for $q_w$ in the expressions for the prices and simplifying the expressions we get for $i=1, 2$
\begin{align*}
p_i & = \frac{p_{\max}}{1+W p_{\max}} (1-q_{-1}) - q_i \left( \frac{W p_{\max}}{1+W p_{\max}}+p_{\max} + \frac{1}{C_i}\right)
\end{align*}
Now the revenue of the providers for $i=1, 2$ is given by
\begin{align*}
R_i(q_i,q_{-i}) & = \frac{p_{\max}}{1+W p_{\max}} (1-q_{-i})q_i \\
& \qquad - q_i^2 \left( \frac{W p_{\max}}{1+W p_{\max}}+p_{\max} + \frac{1}{C_i}\right)
\end{align*}
The best response of each provider $i=1, 2$ is given by
\begin{align*}
\hat{q}_i(q_{-i}) & = \frac{p_{\max}}{2} \frac{1-q_{-i}}{W p_{\max} + \left( p_{\max} + \frac{1}{C_i} \right) (1+W p_{\max})}
\end{align*}
Define the following
\begin{align*}
D & := 4 p_{\max} \sum_{i=1}^2 \frac{1}{C_i} +4W p_{\max}^2 \left( 2+ 2p_{\max} + \sum_{i=1}^2 \frac{1}{C_i}\right)\\
& \quad + 3 p_{\max}^2 +4 \prod_{i=1}^2 \left(\frac{1}{C_i} + W p_{\max} \left( 1+ p_{\max} + \frac{1}{C_i}\right)\right).
\end{align*}
Then the unique fixed point for $i=1, 2$ is given by
\begin{align*}
q_i^* & = \frac{2 \frac{p_{\max}}{C_{-i}} + p_{\max}^2 +2 W p_{\max}^2 \left( 1+ p_{\max} + \frac{1}{C_{-i}}\right)}{D}
\end{align*}
Using these expressions we get the following
\begin{align*}
& q_w^*  = \frac{W p_{\max}}{1+W p_{\max}} \\
& \qquad \times \frac{\prod_{i=1}^2\left(p_{\max} +\frac{2}{C_i}+2 W p_{\max} \left(1+p_{\max} +\frac{1}{C_i} \right)\right)}{D} \\
& \forall i=1, 2 \;\; \frac{p_i^*}{p_{\max}}  =  \\
& \qquad 
\frac{\left( 2 W p_{\max} + W p_{\max}^2 +(1+W p_{\max}) \left(p_{\max} + \frac{1}{C_i}\right)\right)}{(1+W p_{\max}) }  \\
& \qquad \times \frac{\left( 2 W p_{\max} + W p_{\max}^2 +(1+W p_{\max}) \left(p_{\max} + \frac{2}{C_{-i}}\right)\right)}{D}
\end{align*}
It is easy to show that $(p_1^*,p_2^*)$ is a Nash equilibrium. The figures of merit are then given by
\begin{align*}
R_1(q_1^*,q_2^*) & =p_1^* q_1^*, \quad R_2(q_2^*,q_1^*)  =p_2^* q_2^*,\\
S_{ow}(q_1^*,q_2^*) & = \frac{p_{\max}}{2} (q_1^*+q_2^*+q_w^*)^2, \\
T_{ow}(q_1^*,q_2^*) & = R_1(q_1^*,q_2^*) + R_2(q_2^*,q_1^*) + S(q_1^*,q_2^*)
\end{align*}

\section{Perfect competition}\label{sec:competition}

Under the perfect competition (or no market power) assumption, the revenue maximization problem is a convex problem with the unique maximizer given by $q_c(x)$ such that
\begin{align*}
\lambda =  l_c\left(\frac{q_c(x)}{C(x)} \right) + \frac{q_c(x)}{C(x)} l_c^\prime\left(\frac{q_c(x)}{C(x)} \right)
\end{align*}
Comparing with the expression, this implies that the optimal price is given by
\begin{align*}
p(x) = \frac{q_c(x)}{C(x)} l_c^\prime\left(\frac{q_c(x)}{C(x)} \right)
\end{align*}
Additionally, it also follows that $\tfrac{q_c(x)}{C(x)} \equiv \bar{\alpha}$, i.e., independent of $x$, since $\alpha l_c^\prime(\alpha) + l_c(\alpha)$ is an increasing function of $\alpha$. Now $\bar{\alpha}$ is given by the unique solution in $[0, 1/C]$ of
\begin{align*}
P(\alpha C) = l_c(\alpha) + \alpha  l_c^\prime(\alpha) 
\end{align*}
It is easily discerned that $\bar{\alpha} \in (0,1/C)$. The price is given by $p(x) \equiv \bar{\alpha}  l_c^\prime(\bar{\alpha})$. 

The total revenue of all the providers is given by 
\begin{align*}
R_c=\int_0^x R_c(x) dx = \bar{\alpha}^2  l_c^\prime(\bar{\alpha}) C
\end{align*}
The consumer surplus is given by
\begin{align*}
S_c & = \int_0^{\int_0^1 q_c(x) dx} P(q) dq - P\left(\int_0^1 q_c(x) dx\right) \int_0^1 q_c(x) dx\\
& = \int_0^{\bar{\alpha} C} P(q) dq - \bar{\alpha} P(\bar{\alpha} C) C \\
& = \int_0^{\bar{\alpha} C} P(q) dq - \bar{\alpha} l_c(\bar{\alpha}) C - \bar{\alpha}^2  l_c^\prime(\bar{\alpha}) C
\end{align*}
Knowing these two, the total welfare is 
\begin{align*}
T_c = R_c+S_c = \int_0^{\bar{\alpha} C} P(q) dq - \bar{\alpha} l_c(\bar{\alpha}) C
\end{align*}
The partial derivative of $\bar{\alpha}$ in $C$ is given by
\begin{align*}
\frac{\partial \bar{\alpha}}{\partial C} = \frac{\bar{\alpha} P^\prime(\bar{\alpha} C)}{2 l_c^\prime(\bar{\alpha}) + \bar{\alpha} l_c^{\prime\prime}(\bar{\alpha}) - C P^\prime(\bar{\alpha} C) }
\end{align*}
which is negative. From this the partial derivative of $\int_0^1 q_c(x) dx = \bar{\alpha} C$ is given by
\begin{align*}
\frac{\partial \int_0^1 q_c(x) dx}{\partial C}  & = C \frac{\partial \bar{\alpha}}{\partial C} + \bar{\alpha} \\
& = \bar{\alpha} \frac{2 l_c^\prime(\bar{\alpha}) + \bar{\alpha} l_c^{\prime\prime}(\bar{\alpha})}{2 l_c^\prime(\bar{\alpha}) + \bar{\alpha} l_c^{\prime\prime}(\bar{\alpha}) - C P^\prime(\bar{\alpha} C) }
\end{align*}
which is positive. Using all of this we have
\begin{align*}
\frac{\partial T_c}{\partial C} & = \left( C \frac{\partial \bar{\alpha}}{\partial C} + \bar{\alpha}\right) P(\bar{\alpha} C) - \bar{\alpha} l_c(\bar{\alpha}) - C \frac{\partial \bar{\alpha}}{\partial C} ( \bar{\alpha} l_c^\prime(\bar{\alpha}) + l_c(\bar{\alpha}) ) \\
& = C  \frac{\partial \bar{\alpha}}{\partial C}  ( P(\bar{\alpha} C) - \bar{\alpha} l_c^\prime(\bar{\alpha}) - l_c(\bar{\alpha}) ) + \bar{\alpha} ( P(\bar{\alpha} C) - l_c(\bar{\alpha}) ) \\
& = \bar{\alpha}^2 l_c^\prime(\bar{\alpha})
\end{align*}

Finally, we will show that the particular equilibrium that we consider maximizes total welfare so that we get an efficient allocation. The total revenue, consumer surplus and total welfare for a valid traffic profile $q_c(x)$ (equivalent to asking price $p(x)$ when no provider demands an excessive price) are given by
\begin{align*}
& R_c  = \int_0^1 p(x) q_c(x) dx \\
& = P\left( \int_0^1 q_c(u) du \right)  \left( \int_0^1 q_c(u) du\right) -  \int_0^1 q_c(x) l_c\left(\frac{q_c(x)}{C(x)} \right)   dx \\
& S_c  = \int_0^{ \int_0^1 q_c(u) du} P(q) dq - P\left( \int_0^1 q_c(u) du \right)  \left( \int_0^1 q_c(u) du\right) \\
& T_c  = \int_0^{ \int_0^1 q_c(u) du} P(q) dq - \int_0^1 q_c(x) l_c\left(\frac{q_c(x)}{C(x)} \right)   dx
\end{align*}
Since $P(q)$ is an decreasing function of $q$ and $\alpha l_c^\prime(\alpha) + l_c(\alpha)$ is an increasing function of $\alpha$, it is easily verified that the total welfare $T$ is a concave functional of the traffic profile $q_c(\cdot)$. The Frechet differential of $T$ in direction $h(\cdot)$ at $q_c(\cdot)$ is 
\begin{align*}
\delta T_c(q_c; h)& = P\left(\int_0^1 q_c(x) dx\right) \int_0^1 h(x) dx -\\
&  \quad \int_0^1 \left(l_c\left(\frac{q_c(x)}{C(x)}\right) + \frac{q_c(x)}{C(x)}l_c^\prime\left(\frac{q_c(x)}{C(x)}\right) \right) h(x) dx.
\end{align*} 
The constraints on $q_c(\cdot)$ are $\int_0^1 q_c(x) dx \in [0, q_{\max}]$, where, without loss of generality, we assume $q_{\max}=1$, and $q_c(x) > 0$ for all $x\in [0,1]$. Note that this is a convex constraint. Therefore, using \cite[Theorem 2, Chapter 7]{LuenbergerBook1968},
we find that the choice of $q_c(x) = \bar{\alpha} C(x)$ (an interior point of the constraint set) also maximizes the total welfare as the Frechet differential is $0$ in all feasible directions, i.e., it is the efficient allocation as well.

\section{Perfect competition with whitespace}\label{sec:competitionplus}

We now generalize the perfect competition model by adding $W$ bandwidth as whitespace. The user equilibrium now becomes
\begin{align*}
\lambda & = P\left( q_w + \int_0^1 q_c(y) dy \right) \\
& =  l_w\left(\frac{q_w}{W} \right) = p(x) + l_c\left(\frac{q_c(x)}{C(x)} \right)  \quad \forall x \in [0,1] 
\end{align*}
Using the same logic as before, the revenue maximizing price $p(x) =  \tfrac{q_c(x)}{C(x)} l_c^\prime\left(\tfrac{q_c(x)}{C(x)} \right)$ and $\tfrac{q_c(x)}{C(x)}\equiv\alpha_w$ where $\alpha_w$ and user demand in white-space $q_w^*$ satisfy
\begin{align*}
l_w\left(\frac{q_w^*}{W} \right) = P(q_w^* + \alpha_w C) = \alpha_w l_c^\prime(\alpha_w) + l_c(\alpha_w) 
\end{align*}
Solving for $q_w^*$ in terms of $\alpha_w$ we get
\begin{align*}
q_w^* = D\big(\alpha l_c^\prime(\alpha) + l_c(\alpha)\big)-\alpha C
\end{align*}
Using this to eliminate $q_w^*$, $\alpha_w$ must solve
\begin{align*}
l_w\left( \frac{D\big(\alpha l_c^\prime(\alpha) + l_c(\alpha)\big)-\alpha C}{W}\right) = \alpha l_c^\prime(\alpha) + l_c(\alpha)
\end{align*}
It is easy to see that the LHS is a decreasing function of $\alpha$ and the RHS is an increasing function of $\alpha$ such that they cross each other in $[0, \bar{\alpha}]$. This implies that there is a unique solution $\alpha_w \in (0, \bar{\alpha})$. 

The total revenue of the providers, consumer surplus and total welfare are given by
\begin{align*}
R_{cw} & = \alpha_w^2 l_w^{\prime}(\alpha_w) C \\
S_{cw} & = \int_0^{\alpha_w C + q_w^* } P(q) dq - \big( \alpha_w l_c^\prime(\alpha_w) + l_c(\alpha_w)  \big) \big(\alpha_w C + q_w^*\big) \\
T_{cw} & = \int_0^{\alpha_w C + q_w^* } P(q) dq - \big( \alpha_w l_c^\prime(\alpha_w) + l_c(\alpha_w)  \big) q_w^* \\
& \qquad - l_c(\alpha_w) \alpha_w C 
\end{align*}

\section{Justification of perfect competition model}\label{sec:perfectcompetition}

We will mathematically justify the perfect competition model as a limiting case of many providers. For this we will restrict attention to the symmetric case and consider a symmetric Nash equilibrium. Assume that there are $n$ providers each with capacity $C/n$. Let provider $i\in \{1,\dotsc,n\}$ demand price $p_i$ and get a share $q_c(i)$ of the traffic. We will assume, without loss of generality, that no provider asks a price so high that no traffic gets routed to it. Then the user equilibrium condition is
\begin{align*}
P\left( \sum_{j=1}^n q_c(j) \right) = p_i + l_c\left( \frac{q_c(i) n}{C}\right) \qquad \forall i\in \{1,\dotsc,n\}
\end{align*}
From this we have
\begin{align*}
\forall i\in \{1,\dotsc,n\} \qquad p_i = P\left( \sum_{j=1}^n q_c(j) \right) - l_c\left( \frac{q_c(i) n}{C}\right)
\end{align*}
so that we can equivalently consider the traffic served for the Nash equilibria. 

Since we will be interested in symmetric equilibria, without loss of generality consider provider $1$. The best response of provider $1$ to actions of the other providers is to choose $q_c(1)$ to maximize its revenue $R(1)$ which is given by
\begin{align*}
R(1)=q_c(1) \left( P\left( q_c(1) + \sum_{j=2}^n q_c(j) \right) - l_c\left( \frac{q_c(1) n}{C}\right)\right)
\end{align*}
It is easy to verify that the objective is concave in $q_c(1)$. The partial derivative of the revenue in $q_c(1)$ is given by 
\begin{align*}
& \frac{\partial R(1)}{\partial q_c(1)}  =  \\
& \quad P\left( q_c(1) + \sum_{j=2}^n q_c(j) \right) + q_c(1) P^\prime\left( q_c(1) + \sum_{j=2}^n q_c(j) \right) \\
& \quad - \frac{q_c(1) n}{C} l_c^\prime\left( \frac{q_c(1) n}{C}\right) - l_c\left( \frac{q_c(1) n}{C}\right)
\end{align*}
If $\sum_{j=2}^n q_c(j) < 1$, then there exists a unique solution in $\left(0, 1- \sum_{j=2}^n q_c(j) \right)$. Therefore, the unique symmetric equilibrium $q_c^n$ solves
\begin{align*}
P\left( n q_c \right) + q_c P^\prime\left( n q_c\right) = \frac{q_c n}{C} l_c^\prime\left( \frac{q_c n}{C}\right) + l_c\left( \frac{q_c n}{C}\right)
\end{align*}
with $n q_c^n \in (0, 1)$. 

Now we will consider the limiting behavior of $n$ increasing without bound. Since $n q_c^n$ is bounded, along subsequences we have a limit. Let $\{n_k\}_{k=1}^{\infty}$ be one such subsequence and let $q_c^k$ be the limiting value of $n_k q_c^{n_k}$. Then $\lim_{k\rightarrow\infty}q_c^{n_k} = 0$, $q_c^k\in[0,1]$ and $q_c^k$ satisfies
\begin{align*}
P(q_c) = \frac{q_c}{C} l_c^\prime\left( \frac{q_c}{C}\right) + l_c\left( \frac{q_c}{C}\right)
\end{align*}
It can be verified that the equation above has exactly one solution in $(0,1)$, say $q_c^*$. Therefore, $\lim_{n\rightarrow\infty} n q_c^n = q_c^*$. Note that $q_c^*$ is exactly what results from our perfect competition analysis, see \eqref{eq:pceq}.

\section{Proof of Lemma}\label{sec:concave}

Without loss of generality consider provider $1$. Note the following:
\begin{align*}
q_1^* & =  \frac{p_{\max}^2 + 2 \frac{p_{\max}}{C_2}}{3 p_{\max}^2 + 4 \frac{p_{\max}}{C_2}} \frac{C_1 \left( 3 p_{\max}^2 + 4 \frac{p_{\max}}{C_2}\right)}{C_1 \left( 3 p_{\max}^2 + 4 \frac{p_{\max}}{C_2}\right)+ 4 p_{\max}  + \frac{4}{C_2}} \\
p_1^* & = p_{\max} \frac{p_{\max}^2 + 2 \frac{p_{\max}}{C_2}}{3 p_{\max}^2 + 4 \frac{p_{\max}}{C_2}}  \\
& \quad \times
\frac{C_1 \left( 3 p_{\max}^2 + 4 \frac{p_{\max}}{C_2}\right)+ 4 p_{\max}  + \frac{4}{C_2}-p_{\max}}{C_1 \left( 3 p_{\max}^2 + 4 \frac{p_{\max}}{C_2}\right)+ 4 p_{\max}  + \frac{4}{C_2}}
\end{align*}
We are interested in the properties of $R_1(p_1^*,p_2^*)=p_1^* q_1^*$ as a function of $C_1$. From the above, we find that
\begin{align*}
R_1(p_1^*,p_2^*) \propto \frac{a C_1}{a C_1 + b}\frac{a C_1+b d}{a C_1 + b} = 1-\frac{b (2-d)}{a C_1 +b}+ \frac{b^2(1-d)}{\left( a C_1 +b\right)^2},
\end{align*}
where 
\begin{align*}
a & = 3 p_{\max}^2 + 4 \frac{p_{\max}}{C_2} > 0 \\
b & = 4 p_{\max}  + \frac{4}{C_2} > 0, 
\end{align*}
and $d \in (3/4,1]$ (limits obtained by varying $C_2$). Thus, the first and second partial derivative of $R_1$ in $C_1$ are given by
\begin{align*}
\frac{\partial R_1}{\partial C_1} & \propto  (2-d) a C_1 + b d\\
\frac{\partial^2 R_1}{\partial C_1^2} & \propto - \Big( b (2d-1) + a (2-d) C_1 \Big),
\end{align*}
with the constant of proportionality for the first derivative given by
\begin{align*}
p_{\max} \left( \frac{p_{\max} +  \frac{2}{C_2}}{3 p_{\max} + \frac{4}{C_2}} \right)^2 \frac{a b }{\left( a C_1 + b\right)^3}.
\end{align*}
Since $d \in (3/4,1]$, it follows that the second derivative of $R_1$ is negative proving that $R_1$ is a strictly concave function of $C_1$.

\bibliographystyle{IEEEtran}

\end{document}